\documentclass[english,aps,prl,superscriptaddress]{revtex4}
\usepackage{hyperref}
\usepackage{graphicx}
\hypersetup{colorlinks=true,linkcolor=blue,citecolor=red,urlcolor=black}
\begin{document}
\title{Broadband Tuning of Optomechanical Cavities}

\author{Gustavo S. Wiederhecker}
\affiliation{School of Electrical and Computer Engineering, Cornell
University, Ithaca, NY, 14853}
\author{Sasikanth Manipatruni}
\affiliation{School of Electrical and Computer Engineering, Cornell
University, Ithaca, NY, 14853}
\author{Sunwoo Lee}
\affiliation{School of Electrical and Computer Engineering, Cornell
University, Ithaca, NY, 14853}
\author{Michal Lipson}

\affiliation{School of Electrical and Computer Engineering, Cornell
University, Ithaca, NY, 14853}
\affiliation{Kavli Institute for Nanoscale Science at Cornell, Ithaca,
NY, 14853}


\date{\today}
\begin{abstract} 
We demonstrate broadband tuning of an optomechanical microcavity optical
resonance by exploring the large optomechanical coupling of a
double-wheel microcavity and its uniquely low mechanical stiffness. Using a
pump laser with only 13 mW at telecom wavelengths we show tuning of the silicon nitride microcavity resonances over 32 nm. This corresponds to a tuning power efficiency of only 400 $\mu$W/nm. By choosing a
relatively low optical Q resonance ($\approx$18,000) we prevent the
cavity from reaching the regime of regenerative optomechanical oscillations.  The static mechanical displacement induced by optical gradient
forces is estimated to be as large as 60 nm.
\end{abstract}

\maketitle


\section{Introduction}

The control of light using optical microcavities has important
applications spanning from quantum aspects of light-matter interaction
\cite{GroHamVan09,VahHerKnu09,SchRivAne08,HanSch75} to information routing in advanced photonic networks \cite{LeeBibShe09,RosLinPai09,LirManLip09,HanMicDaq08,BibSheLee08,SheWanChe08}. The
key characteristic of these cavities is their resonant response, which
occurs only when the wavelength of light is an integer fraction of the
cavity's optical path length. Therefore in order to reconfigure these
devices, a method to tune the optical path length is necessary.
Electro-optic tuning has been demonstrated using LiNBO$_3$ \cite{WanChuLin07,GuaPobRez07} however
only sub-nm tuning was achieved; thermo-optic \cite{RejChaFuw07,BibSheLee08} or free-carrier
injection based \cite{LirManLip09,LeeBibShe09} tuning has also been demonstrated with tens of nm
tuning range. These methods however not only are limited to materials
with high thermo-optic coefficient or strong free-carrier dispersion,
but also require high temperatures ($>$ 400 K) or suffer from
free-carrier induced losses \cite{LeeBibShe09,LirManLip09,RejChaFuw07}. Another way to control the
cavity length is to manipulate their mechanical degrees of freedom \cite{RosLinPai09,WieCheGon09,FraDeoMcC10,TakKanKok08,YaoLeuLee07,EicMicPer07,HuaZhoCha0803,RieMauHal0410,AlePerPai1004,PerCohMee10}, such manipulation can be achieved using the optical forces
provided by photons circulating inside the optical microcavities \cite{RosLinPai09,WieCheGon09,EicMicPer07,VanRoe10,LiPerXio08}. For example, it has been recently shown that optical gradient
forces can be used to actuate the mechanical motion in these cavities
with tuning ranges exceeding 2 nm \cite{RosLinPai09,WieCheGon09,VanRoe10}, and also proposed as a tuning method for various waveguide and microcavity parameters \cite{MaPov0909,RakPopSol0711}. 
Here we demonstrate that the use of optical forces can provide tuning of a microcavity resonance over 30 nm using only 13 mW of laser power.
This device power efficiency is 400 $\mu$W/nm and its tuning range can
span the full telecom C or L-band. 
\begin{figure*}[ht] \centering\includegraphics[width=141.8mm]{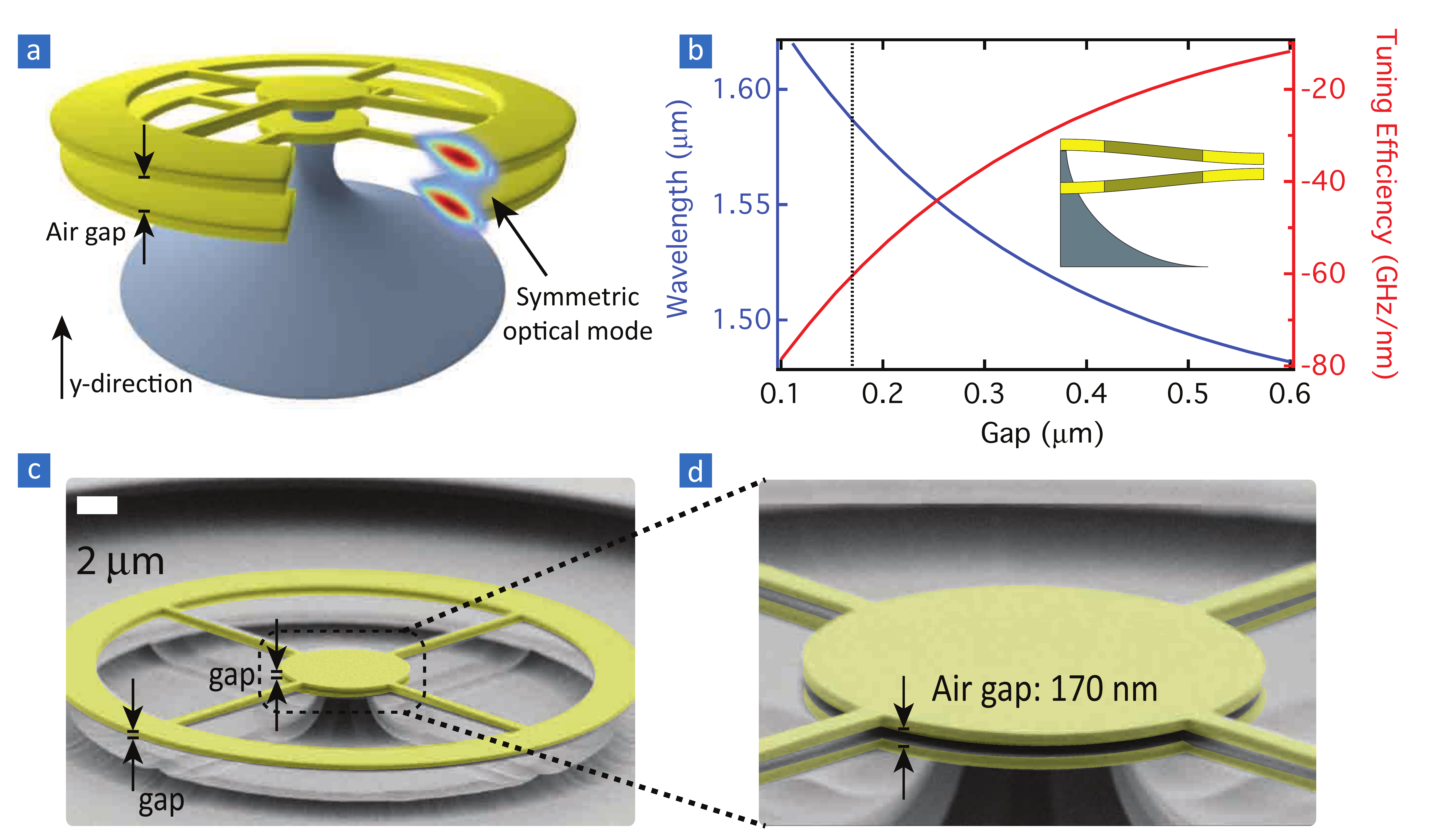}
\caption{\label{Fig1}Optical force actuated optical microcavity. a. Schematic of the
device with a sliced cross-section showing the TE$_1$ symmetric optical
mode profile. b.  Optical resonant wavelength (blue curve, left axis)
and optomechanical tuning efficiency (red curve, right axis) dependence
with the air gap between the rings. The black-dashed line indicates a
gap of 170 nm, close to the fabricated cavity. c,d. Scanning electron
micrograph of two vertically stacked ring cavites.} \end{figure*}

\section{Optical forces in double-ring cavities}

To achieve a strong per-photon optical gradient force we choose to work
with a coupled pair of microring cavities similar to our previous work \cite{WieCheGon09}. As illustrated in the schematic of Fig. \ref{Fig1}a, it consists of two vertically stacked microring cavities. The subwavelength air-gap between the rings allows for strong coupling between the optical modes of the two rings. For air gaps around 170 nm, the optical modes of each ring evanescently couple to each other forming symmetric and anti-symmetric optical super-modes; in Fig. \ref {Fig1}a we show the electric field profile of the symmetric one. The coupling induces a splitting in their resonant wavelength which depends exponentially on the gap between the rings, as shown in the blue curve of Fig. \ref{Fig1}b for the symmetric optical super-mode. Therefore one can use this gap-dependence of the optical resonances to tune the microcavity's optical response. According to the simulated curve in Fig. \ref{Fig1}b a telecom band tuning of 30 nm can be achieved with a gap change of only 60 nm. In this optical microcavity, the force to drive such a change can be derived from the optical field gradient in the cavity. The optical energy inside the cavity depends on
the optical mode  resonant frequency ($U=N\hbar \omega $ where is $N$ is the number of photons circulating in the cavity  and $\omega$ is the cavity mode optical resonant frequency), therefore the cavity's optical energy also depends on the gap between the two ring resonators; an optical force between these rings should follow using a virtual work approach \cite{RakPopSol0711,PovLonIba05}. The optical potential energy change inside the cavity ($ N \hbar \delta \omega$) must correspond to the mechanical work realized on the microrings ($-F\delta y$), therefore the optical force is given by $F=-N\hbar g_{om} $, where  $g_{om}=\partial \omega /\partial y$ is the optomechanical tuning efficiency. This tuning efficiency is shown as the red curve in Fig. \ref{Fig1}b. When operating at small gaps ($\approx$170 nm) the
resonant frequency tuning efficiency $g_{om} /2\pi $ can be as high as
60 GHz/nm, this corresponds to a 40 fN/photon optical gradient force.

\begin{figure}[ht] \centering\includegraphics[width=141.8mm]{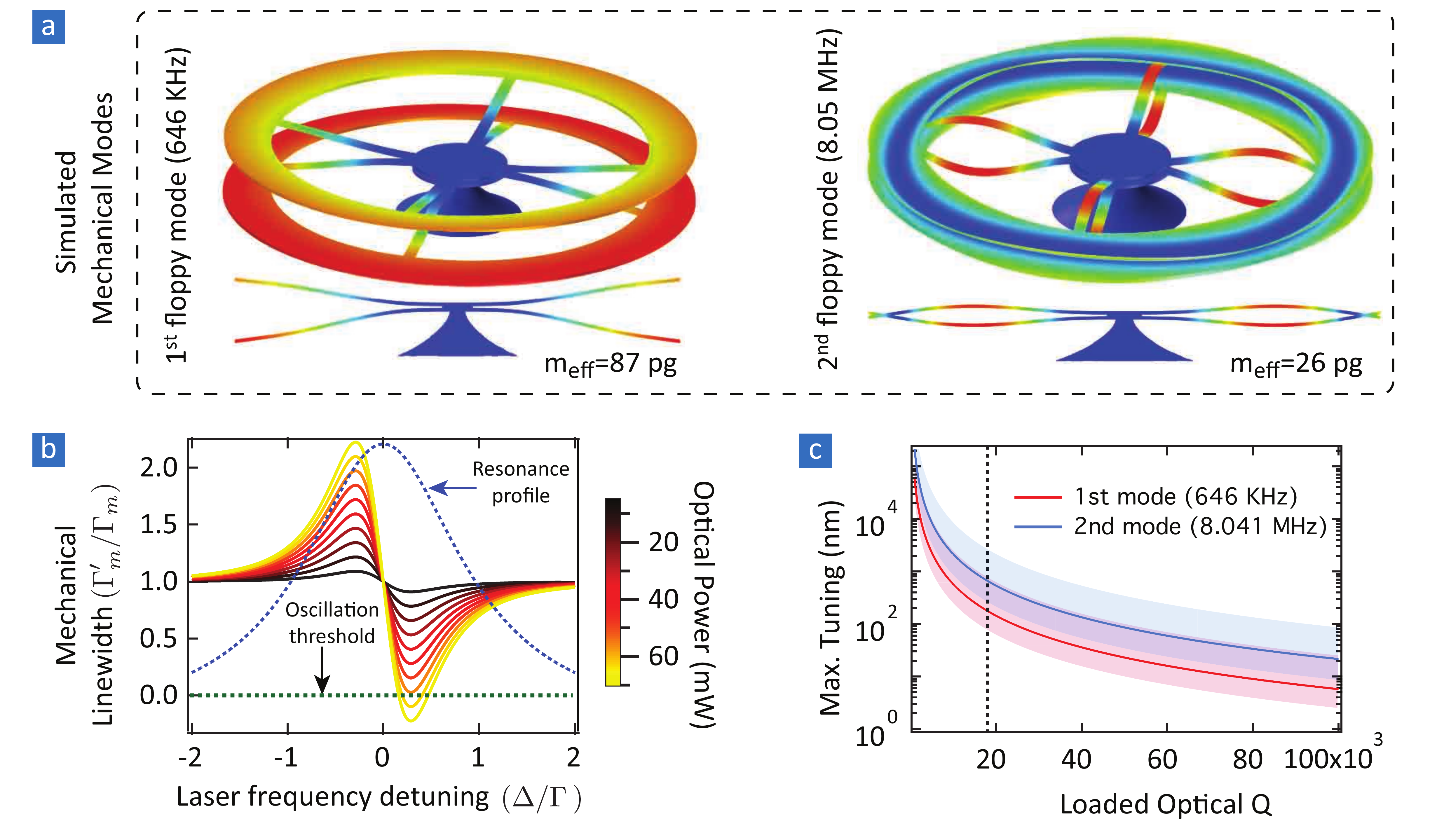}
\caption{\label{Fig2}Limit for static tunability of optomechanical cavities. (a) Two simulated mechanical floppy modes and their respective effective masses. (b)
Effective mechanical linewidth normalized by the intrinsic mechanical
linewidth ($\Gamma '_{m} /\Gamma _{m} $) as a function of the
normalized pump laser frequency detuning ($\Delta/\Gamma$). The blue-detuned pump laser induces gain, which above a
certain threshold induces regenerative mechanical oscillations. The
dashed-blue line indicates the optical resonance profile whereas the
dashed-green line shows the oscillation threshold. (c) Maximum static
tuning predicted by Eq. \ref{Eq2} before reaching oscillation threshold versus the loaded optical quality factor. The different lines corresponds to the threshold for the two different mechanical modes
shown in part (a). The dashed vertical line indicates the loaded optical $Q$ of the tested device.} \end{figure}

The fabricated microcavity structure is optimized to achieve maximum optical frequency tuning and per-photon optical gradient force. Such optomechanical frequency tuning can be expressed as a function of the optical power in the cavity. When a pump laser excites the symmetric super-mode resonance, the flexible spokes bend in response to the optical gradient force. The gap between the rings reduces and the optical resonant frequency of this mode decreases as shown in Fig. \ref{Fig1}b. The cavity optical frequency shift due to mechanical
displacement is simply given by $\Delta \omega =g_{om} \Delta y$, where
$\Delta y$ is the gap change between the rings. These relations can be
combined with Hooke's law ($F=k\Delta y/2$, $k$ is the stiffness of the rings in response to the gradient optical force\cite{comment1}) to yield an expression
for the cavity frequency shift and the optical power coupled to the
cavity,
\begin{equation} \label{Eq1} \Delta \omega =-\frac{2Q_{i}
g_{om}^{2} }{\omega _{0}^{2} k} P_{d} , \end{equation}
where $Q_{i} $ is the intrinsic optical quality, $\omega_0$ is the optical resonant frequency, and $P_{d} =(1-T)P_{in} $ is the power dropped to the cavity resonance with normalized transmission $T$. According to Eq. \ref{Eq1}, the optomechanical tuning $g_{om} $ and the beam
stiffness $k$ are the two cavity parameters that can be engineered in
order to achieve large optical frequency tuning. There are however
practical limits to them; to increase $g_{om} $, as shown in Fig. \ref{Fig1}b, it
is necessary to have cavities with small gaps. When the gaps are smaller
than 100 nm however, the fabrication yield is considerably smaller since
the structures tend to collapse because of the short-range Van der Wals
interaction \cite{DelDe-Kna05}.  This also limits the smallest spring constant that
can be practically achieved.

The double-ring optical cavity used here  can have optical and mechanical parameters, such as $g_{om}$, $Q_i$ and $k$ that increases the threshold for regenerative mechanical  in the cavity \cite{SchRivAne08,LinRosJia09,CarRokYan05} and enables large static tuning. When the regenerative oscillation threshold is reached, one or more mechanical modes of the structure will oscillate with a large amplitude leading to a strong modulation of the light transmitted by the cavity. In Fig. \ref{Fig2}a we show the mechanical displacement profile of two floppy mechanical modes that will be driven by the optical gradient force. The threshold optical power at which regenerative mechanical
oscillations begins will define the maximum static frequency tuning, i.e., an upper limit for the static operation of our device. As we illustrate in Fig. \ref{Fig2}b, when the pump laser is blue-detuned
with respect to the cavity frequency ($\Delta \equiv \omega _{p} -\omega_{0} >0$), it provides optical gain for the thermally excited mechanical modes and effectively reduces the mechanical damping, narrowing the mechanical resonance linewidth ($\Gamma '_{m}$). At the threshold power $P_{th} $, the optomechanical gain exceeds
the intrinsic losses of the mechanical modes ($\Gamma '_{m} <0$, dashed black line in
Fig. \ref{Fig2}b) and they enter into regenerative oscillations \cite{WieCheGon09,LinRosJia09,SchArcRiv09,SchRivAne08}. As a result, any optical signal going through the cavity is strongly modulated at the mechanical frequency $\Omega_m$ of these modes. The floppy mechanical modes shown in Fig. \ref{Fig2}a are the first two mechanical modes that are strongly driven by optical field due to their mostly vertical and opposing (or bright) motion of the two rings. The threshold input power to achieve regenerative oscillations for a mechanical mode with effective motional mass $m_{eff}^{(m)}$,  optomechanical coupling rate $g_{om}^{(m)}$, and mechanical quality factor $Q_m=\Omega_m/\Gamma_m$ is given by $P^{(m)}_{th} =m_{eff}^{(m)}\Omega_m \omega_{0}^{4}(8 Q_{m} Q^{3}(g_{om}^{(m)})^2\eta_c )^{-1} $, where $\eta_c\equiv(1\pm\sqrt{T_{min}})/2$ is the ideality coupling factor for an undercoupled $(-)$ or overcoupled $(+)$ cavity, $T_{min}$ is the transmission value exactly on resonance, and $Q=Q_i(1-\eta_c)$ is the loaded optical quality factor \cite {SchRivAne08,LinRosJia09,SchArcRiv09,SpiKipPai0307,AneRivSch08,PinHadHei99}. Here we assumed that the cavity is excited close to the optimal cavity frequency detuning $\Delta\approx\ -\Gamma/2$ (where $\Gamma=\omega_0/Q$) and that the cavity parameters are within the unresolved sideband limit, $\Omega_{m} \ll \Gamma $. At this detuning point, the cavity transmission is given by $T(\Delta=-\Gamma/2)=1+2(\eta_c-1)\eta_c$. Using this transmission value and the power threshold expression above together with Eq. \ref{Eq1}, an expression can be derived for the maximum static frequency shift for an optomechanical cavity as limited by optomechanical oscillations of the $m^{th}$ mechanical mode, 
\begin{equation}
\label{Eq2} \Delta \omega^{(m)}_{th} = -\frac{m_{eff}^{(m)}\Omega_m }{2 k Q_{m} } \left(\frac{g_{om}}{g_{om}^{(m)}}\right)^2 \left(\frac{\omega_0}{Q}  \right)^2\approx  -\frac{m_{eff}^{(m)}\Omega_m }{2 k Q_{m} } \left(\frac{\omega_0}{Q}  \right)^2.
\end{equation}
In the case of the fundamental anti-symmetric (bright) mechanical mode ($m=1$), this expression does not depend on the optomechanical tuning efficiency $g_{om}$ since $g_{om}^{(1)}\approx g_{om}$, a high value of $g_{om}$ however ensures that large tuning can be achieved using low optical powers (see Eq. \ref{Eq1}). For double-ring cavities however many mechanical modes will have similar $g_{om}^{(m)}$, for example $g_{om}^{(2,3)}/g_{om}^{(1)}\approx (60\%,64\%)$, where (2,3) stands for the second and third order bright mechanical modes, therefore Eq. \ref{Eq2} can still predict the maximum frequency shift as limited by optomechanical oscillations of the higher order mechanical modes. In Fig. \ref{Fig2}c we show the maximum wavelength tuning predicted by Eq. \ref{Eq2} for a double-ring cavity with a loaded optical $ Q= 18 \times10^3$ (vertical dashed line), each curve represent the maximum wavelength tuning as limited by regenerative oscillations from the two mechanical modes shown in Fig. \ref{Fig2}a. The red and blue lines corresponds to the first and second mechanical modes shown in Fig. \ref{Fig2}a with parameters  $\Omega_{m}/2\pi=(0.646,8.041)$ MHz, $m_{eff}=(87, 26)$ pg, $k=1.44$ N/m, and $Q_m=4$. Since these modes may have distinct mechanical quality factors, we represent in the wide blue and red regions the tuning range spanned when the mechanical quality factor vary between $1<Q_m<10$. The overlap of these region show that depending on their mechanical quality factor, the second order mode may reach oscillation threshold before the first mode. These parameters should allow for a maximum static tunability in the few hundred nanometers range as shown in Fig. \ref{Fig2}c.

\section{Experimental Results}

\begin{figure}[ht] \centering\includegraphics[width=141.8mm]{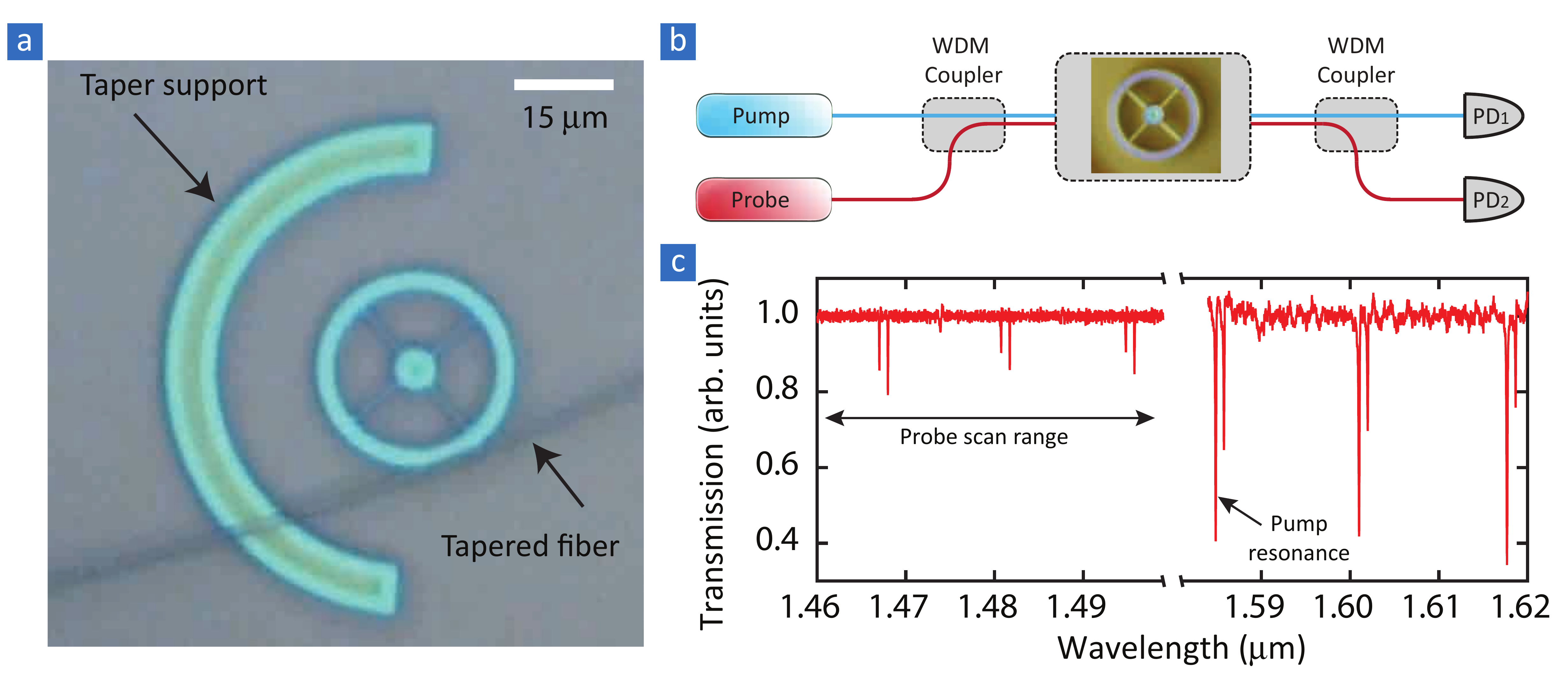}
\caption{\label{Fig3}Experimental setup and cold-transmission. (a) Top view optical
micrograph of the device showing the tapered optical fiber used to
support the device. This is due to change in the
interference pattern as the air-gap between the rings changes.  (b)
Schematic of the experimental setup, PD1,2 denotes the two photodiodes
used to record the pump and probe transmission. (c)  Low power (100 nW)
optical transmission of the cavity highlighting both the probe
(1460-1500 nm) and pump (1575-1620 nm) wavelength region.} \end{figure}

We demonstrate experimentally optomechanical tuning exceeding 30 nm
using only 13 mW of laser power, well below the
regenerative oscillations threshold. The fabricated cavity, shown in Fig. \ref{Fig1}c,d, has a 30 $\mu$m diameter and a 3 $\mu$m wide ring. Each ring is made of 190 nm thick stoichiometric LPCVD (low-pressure chemical vapor deposition) Si$_3$N$_4$. The details of the fabrication process can be found elsewhere  \cite{WieCheGon09}. The spokes have cross-section dimensions of 190 x 500 nm. The air gap between the rings is about 170 nm. A top view of the device
under test is shown in Fig. \ref{Fig3}a and the experimental setup schematic is
shown in Fig. \ref{Fig3}b. A typical low power transmission spectrum obtained
using a tapered optical fiber coupled evanescently to the cavity is
shown Fig. \ref{Fig3}c. The loaded optical quality factor of the pump resonance was
$Q=18,000$ whereas the mechanical quality factors was $Q_{m} =2$ for the mechanical mode at 8.05 MHz, for the first order mode at 646 KHz it was too low and was not measured since it was below the noise level in our direct detection setup. Such low
mechanical quality factor is typical in double-ring cavities due to the
strong damping caused by gas trapped between the rings \cite{WieCheGon09,LinRosJia09,BaoYan07,RosLinPai09}.

\begin{figure}[htbp] \centering\includegraphics[width=141.8mm]{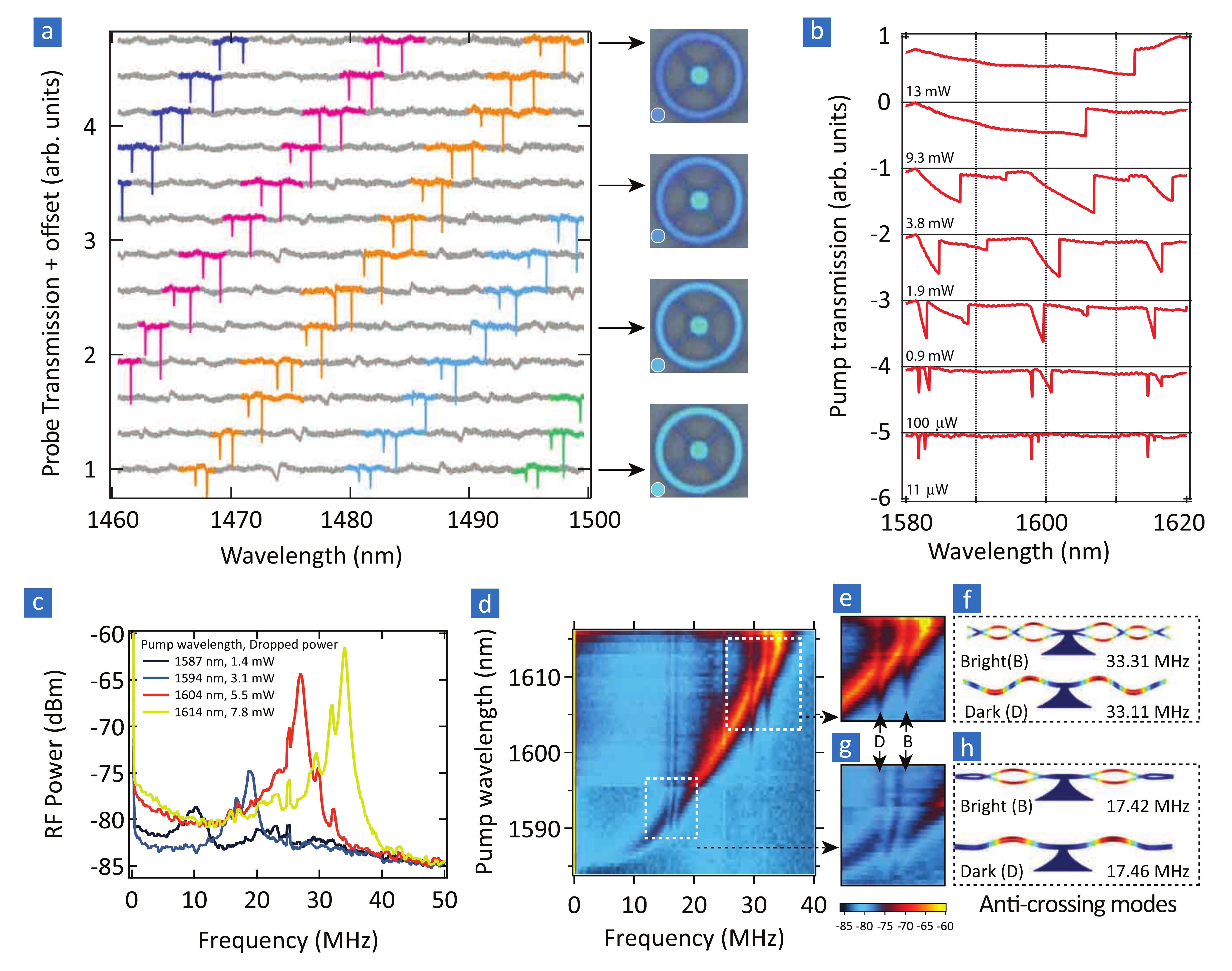}
\caption{\label{Fig4}Optomechanical tuning of double-ring cavity. (a) Measured probe laser transmission for a pump power of 13 mW. The
different curves are recorded at distinct pump laser detuning from the
cavity resonance, the bottom and top curves are recorder when the pump laser is
out of resonance and fully resonant, respectively; the micrographs on the right show the cavity color recorded corresponding to the transmission curves indicated by the arrows. The embedded movie shows the ring color changing as the optical force builds up on the device.  (b) Measured optical transmission of the pump laser at increasing power levels. (c) RF spectrum
showing the optomechanical amplification of the mechanical resonance,
even at maximum amplification (yellow curve) the measured mechanical
quality factor is 30. (d) RF spectrum of the transmitted pump laser showing
the optical spring effect on the mechanical resonance. The highlighted regions (e,g) show the anti-crossing between the mechanical resonant modes. The false color scale represents the RF power in dBm. (f,g) Simulated bright and dark mechanical modes corresponding to the anti-crossings observed on (e,g).} \end{figure}

To induce the optical force we use a 13 mW tunable external cavity pump
laser centered on an optical resonance at 1580 nm (highlighted in Fig.
\ref{Fig3}c), the polarization of the pump is adjusted to maximize its coupling to the cavity TE mode. As the pump laser wavelength scans this resonance (0.5 nm steps)
from shorter towards longer wavelengths, more power is dropped to the cavity as the laser approaches the
resonance. At each pump wavelength
step, we record the probe transmission over the range indicated by the
horizontal arrow in Fig. \ref{Fig3}c. We summarize our experimental data demonstrating 32 nm optomechanical tuning in Fig. \ref{Fig4}. The most dramatic effect of the optomechanical tuning can be seen by simply observing under a microscope the colour change of the light reflected off the top of the cavity. As the air-gap between the rings is reduced, the thin-film interference effect allow us to see in real time such a change. We show this effect in the sequence of images on right-side of Fig. \ref{Fig4}a, and also on the attached multimedia file on Fig. \ref{Fig4}. A more quantitative evidence is revealed by the pump laser transmission shown in Fig. \ref{Fig4}b. The offset curves correspond to different power levels as the laser scans from shorter to longer wavelengths. As the pump power increases (from bottom to top) the optical resonances show the typical triangular shape of a bistable
cavity \cite{KipVah07,WieCheGon09,RosLinPai09}. Such bistable transmission curves results from the red-shift induced by the change in the air-gap due to optical forces. At the highest pump power of 13 mW, the bistability extend over 32 nm,
which spans twice the optical free spectral range. This indicates that a very large optomechanical shift is induced. The probe transmission curves at
each pump wavelength are shown as offset curves in Fig. \ref{Fig4}a. Since the
pump power actually dropped to the cavity is related to the pump
transmission curve and input power as $P_{d} =(1-T)P_{in} $, the minimum
pump transmission (around 1617 nm) is $T$=0.4 and therefore the
maximum power dropped to the cavity is $P_d$=7.8 mW. Using Eq.
\ref{Eq1} with this power value and a spring constant of
$k$=1.44 N/m, as obtained from static finite element simulations \cite{comment1}, we
calculate the expected wavelength shift to be $\Delta \lambda $=33 nm,
in good agreement with the measured value. 

As the dropped power in the cavity increases we observe tuning of mechanical resonant frequency (i.e., optical spring effect) as well as a reduction in the mechanical resonance linewidth (i.e., optomechanical amplification). These two effects can be seen in Fig. \ref{Fig4}c and \ref{Fig4}d where we
show a density plot of the transmitted pump RF spectrum measured using a
125 MHz photodetector (PD2 in Fig. \ref{Fig4}a). The observed shift in the mechanical
resonant frequency, from 8 MHz to 32 MHz, corresponds to a
stiffening of the mechanical resonator of $k'/k=(\Omega '/\Omega )^{2}
\approx 16$. The stiffening of the resonator is so large that the mechanical mode exhibit anti-crossings \cite{LinRosCha1004} with higher frequency modes, two of such anti-crossings at 17 and 31 MHz are highlighted in Fig. \ref{Fig4}d. Each of these anti-crossings are actual doublets formed by a symmetric (dark) lower frequency mechanical mode $\Omega_d/2\pi=(16,31)$ MHz, shown in Fig. \ref{Fig4}e,f, and an anti-symmetric (bright) higher frequency mechanical mode $\Omega_b/2\pi=(17,32)$ MHz, shown in Fig. \ref{Fig4}g,h; the mechanical frequencies shown in these figures are obtained from finite element simulations and agree with the measurements within  $<10\%$ difference. This identification is justified since after the anti-crossing with the dark modes, the mechanical mode preserves its frequency and appears as a straight vertical trace in density plot of Fig. \ref{Fig4}d, whereas after the anti-crossing with the bright mode, the mechanical mode keeps increasing its frequency due to the spring effect. The major benefit of this stiffening for static tuning applications is the reduction of
thermal driven vibrations, which scales as $\sqrt{k_{B} T/k'}$  according to the equipartition of energy \cite{RosLinPai09,WieCheGon09,KipVah07}.
The reduction of the mechanical linewidth leads to an increase in the mechanical quality factor, as shown in Fig. \ref{Fig4}c which was measured to increase from $Q_m\approx2$ to $Q_m\approx30$. This confirms that our device is indeed well below the optomechanical oscillations threshold.

The optical absorption inside the cavity raises its temperature and also contributes to the measured shift. There is a thermo-optic contribution arising from
the refractive index change ($\delta n=\alpha \Delta T$, where $\alpha
=4\times 10^{-5} $ K$^{-1}$ for Si$_3$N$_4$) and a thermo-mechanical contribution
due to thermal expansion of the cavity. Although such thermal
expansion could also cause a change in the gap between the rings, it has
been shown that the major thermo-mechanical shift is caused by radial
expansion \cite{RosLinPai09,WieCheGon09}. We verified both numerically and experimentally that
these thermal contributions are negligible. Considering both thermal
contributions we may write $\Delta \omega _{th} =g_{T} \delta T+g_{R}
\delta R$ where $g_{T} =\omega \alpha /n_{g} $ and $g_{R} =\omega /2\pi
R$, one can estimate the radial expansion of the ring using the relation
$\delta R/R=\alpha _{L} \delta T$, where $\alpha _{L} \approx10^{-6}$ K$^{-1}$ is the
SiN thermal expansion coefficient.  To estimate the temperature change
we assume an absorption loss of 0.06 dB/cm \cite{GonLevLip09}, which corresponds to an optical 
absorption quality factor of $Q_{abs} = 5.2\times 10^6$. The total thermal
resistance of the cavity, as calculated through the finite element
method \cite{Comsol}, is $R_{th} =7\times 10^{5} $ K/W. Using these parameters, the
estimated temperature change in the ring is $\delta T=R_{th} P_{heat}
$=23 K, where the heating power is calculated from the intra-cavity
energy $U$ as $P_{heat} =\omega _{0} U/Q_{abs} $\cite{WieCheGon09}. Using the above
relation we calculate the thermal contribution to the frequency shift, the thermal expansion term gives $g_T \delta T/2\pi= -4.5$ GHz, whereas the thermo-optic effect gives $g_T \delta T/2\pi= -99$ GHz. The total thermal shift is therefore $\Delta \omega _{th} /2\pi =-103$ GHz, or equivalently $\Delta \lambda_{th} $=0.86 nm, which corresponds to 3\% of the measured shift. To verify that the thermal contribution is indeed small we tested a device in which the two rings collapsed and were stuck to each other and therefore do not experience the usual optomechanical tuning. Such a small contribution from thermal shift was verified experimentally is in agreement with previous results on double-ring and spider-web cavities where the optomechanical tuning is the dominant effect. Some contribution could also arise from the nonlinear Kerr effect, however due to low finesse of our cavity ($\mathcal{F}\approx180$) and relatively large effective mode area ($A_{eff}\approx 8\times 10^{-13}$ m$^2$), we estimated the Kerr contribution $\Delta\lambda_{Kerr}=\lambda n_2 \mathcal{F} P_d/(\pi n_g A_{eff})$, where $n_2=2.5\times 10^{-15}$ cm$^2$W$^{-1}$ is the nonlinear refractive index of Si$_3$N$_4$ and $n_g=1.8$ is the cavity mode group index, to be below 1 pm and therefore negligible in our device.

\section{Conclusions}
In conclusion, we show efficient (400 $\mu$W/nm), broadband (across C \&L bands) tuning of optical resonances using gradient force actuation of optical devices. We also show that competing effects such as thermo-optic effect and Kerr effect contribute only to a small extent to the overall optical frequency shift. This optomechanical tuning approach is not only competitive with other known tuning methods, but also advantageous since it simplifies the fabrication process by avoiding metal contacts etc. Static tuning beyond the one achieved here should be possible using such devices, although further optimization of the resonator parameter, such as spokes thickness, inter-ring gap, ring width, and perhaps higher laser powers may be necessary. While we focused on the static effects in this work, using appropriate experimental conditions, high tuning efficiency gradient force optomechanical devices may offer a potential for studying dynamic effects of radiation and near field forces. 

\section*{Acknowledgements}

The authors acknowledge Long Chen and Jaime Cardenas for valuable
fabrication help, and also Nicholas Sherwood and George Kakarantzas for
their help in building the fiber taper pulling station. We also
acknowledge partial support by Cornell University's Center for Nanoscale
Systems. This work was supported in part by the National Science
Foundation under grant 00446571. This work was performed in part at the
Cornell Nano-Scale Science and Technology Facility (a member of the
National Nanofabrication Users Network) which is supported by the
National Science Foundation, its users, Cornell University and
Industrial users.

\bibliographystyle{apsrev}
\bibliography{Broadband_Tuning_OpEx}
\end{document}